\documentclass[journal]{IEEEtran}
\ifCLASSINFOpdf
\else
\fi

\usepackage[version=3]{mhchem} 
\usepackage{xcolor, siunitx, booktabs, multirow, subcaption, diffcoeff, graphicx, hyperref}

\DeclareMathOperator*{\argmin}{arg\,min}
\newcommand{\Fxyop}{\mathcal{F}_{xy}}
\newcommand{\iFxyop}{\mathcal{F}_{xy}^{-1}}
\newcommand{\Fxybracket}[1]{\hspace{-0.1em}\left\{\hspace{-0.0em}#1\hspace{-0.0em}\right\}}
\newcommand{\Fxysbracket}[1]{\{#1\}}
\newcommand{\Fxy}[1]{\Fxyop\Fxybracket{#1}}
\newcommand{\iFxy}[1]{\iFxyop\Fxybracket{#1}}
\newcommand{\Fxys}[1]{\Fxyop\Fxysbracket{#1}}
\newcommand{\iFxys}[1]{\iFxyop\Fxysbracket{#1}}
\newcommand{\Loss}{\mathcal{L}}
\newcommand{\Grad}{\mathcal{G}}
\newcommand{\GradL}[1]{\Grad(\Loss, #1)}
\newcommand{\cvarx}{v}
\newcommand{\cvary}{w}

\begin{document}
%
\title{Reflection-mode Multi-slice Fourier Ptychographic Tomography}
%
%
%

\author{Jiabei~Zhu,
        Tongyu~Li,
        Hao~Wang,
        Yi~Shen,
        Guorong~Hu,
        and~Lei~Tian
\thanks{Manuscript received May 9, 2025; revised May 9, 2025. This work is funded by Samsung Global Research Outreach (GRO) program, and National Science Foundation (1846784). The authors thank Boston University Shared Computing Cluster for proving the computational resources.
\emph{Corresponding author: Lei Tian}}
\thanks{Jiabei Zhu is with the Department of Electrical and Computer Engineering,  Boston University, Boston, Massachusetts 02215, USA (e-mail: zjb@bu.edu).}
\thanks{Tongyu Li is with the Department of Electrical and Computer Engineering,  Boston University, Boston, Massachusetts 02215, USA (e-mail: tongyuli@bu.edu).}
\thanks{Hao~Wang is with the Department of Electrical and Computer Engineering,  Boston University, Boston, Massachusetts 02215, USA (e-mail: wanghao6@bu.edu).}
\thanks{Yi~Shen is with the Department of Electrical and Computer Engineering,  Boston University, Boston, Massachusetts 02215, USA (e-mail: yishen@bu.edu).}
\thanks{Guorong~Hu is with the Department of Electrical and Computer Engineering,  Boston University, Boston, Massachusetts 02215, USA (e-mail: grhu@bu.edu).}
\thanks{Lei~Tian is with the Department of Electrical and Computer Engineering and the Department of Biomedical Engineering,  Boston University, Boston, Massachusetts 02215, USA (e-mail: leitian@bu.edu).}
}

%
%

\markboth{Journal of \LaTeX\ Class Files,~Vol.~XX, No.~X, May~2025}%
{Shell \MakeLowercase{\textit{et al.}}: Bare Demo of IEEEtran.cls for IEEE Journals}
%



\maketitle

\begin{abstract}
Diffraction tomography (DT) has been widely explored in transmission-mode configurations, enabling high-resolution, label-free 3D imaging.
However, industrial metrology applications, such as semiconductor inspection, typically involve opaque or highly reflective substrates (e.g., silicon or metal), necessitating a reflection-mode imaging configuration.
In this work, we introduce reflection-mode Multi-Slice Fourier Ptychographic Tomography (rMS-FPT) that achieves high-resolution, volumetric imaging of multi-layered, strongly scattering samples on reflective substrates.
We develop a reflection-mode multi-slice beam propagation method (rMSBP) to model multiple scattering and substrate interactions, enabling precise 3D reconstruction. By incorporating darkfield measurements, rMS-FPT enhances resolution beyond the traditional brightfield limit and provides sub-micrometer lateral resolution while achieving optical sectioning.
We validate rMS-FPT through numerical simulations on a four-layer resolution target and experimental demonstrations using a reflection-mode LED array microscope.
Experiments on a two-layer resolution target and a multi-layer scattering sample confirm the method's effectiveness.
Our optimized implementation enables rapid imaging, covering a \qtyproduct{1.2 x 1.2}{\mm} area in 1.6 seconds, reconstructing over $10^9$ voxels within a $\qty{0.4}{\mm^3}$ volume.
This work represents a significant step in extending DT to reflection-mode configurations, providing a robust and scalable solution for 3D metrology and industrial inspection.
\end{abstract}

\begin{IEEEkeywords}
Diffraction tomography, Fourier ptychography, Multiple scattering, 3D Metrology.
\end{IEEEkeywords}

%
\IEEEpeerreviewmaketitle

\section{Introduction}
%
%
%
%
\IEEEPARstart{D}{iffraction} tomography (DT) has been established as a high-resolution, label-free, and non-destructive method for characterizing 3D refractive index (RI) distributions from angle-diverse measurements.
Traditionally, DT has been widely applied in biomedical imaging~\cite{park2018quantitative} for monitoring cellular dynamics~\cite{kim2014white} and analyzing tissue morphology~\cite{merola2017tomographic} using transmission-mode configurations, where light passes through the sample before detection.
Recently, DT has gained traction in industrial metrology and inspection, particularly for semiconductor manufacturing and other high-precision applications~\cite{kim2016large,kang2023accelerated,aidukas2024high}.
However, when imaging samples on strongly reflective substrates, such as semiconductor wafers or metallic surfaces, transmission-mode imaging becomes impractical.
As a result, DT must operate in reflection mode for these critical applications.
In this configuration, the illumination effectively makes a ``double pass'' through the sample, interacting with the reflective substrate before returning to the detector.
While this round-trip propagation introduces additional complexities, including substrate-induced reflections and multiple-scattering effects, it also enhances sensitivity to subtle phase variations in thin layers.
These factors pose significant challenges for achieving accurate 3D reconstructions, requiring advanced modeling approaches capable of simultaneously handling strong scattering and reflection.

\begin{figure*}[!t]
\centering
\includegraphics[width=\textwidth]{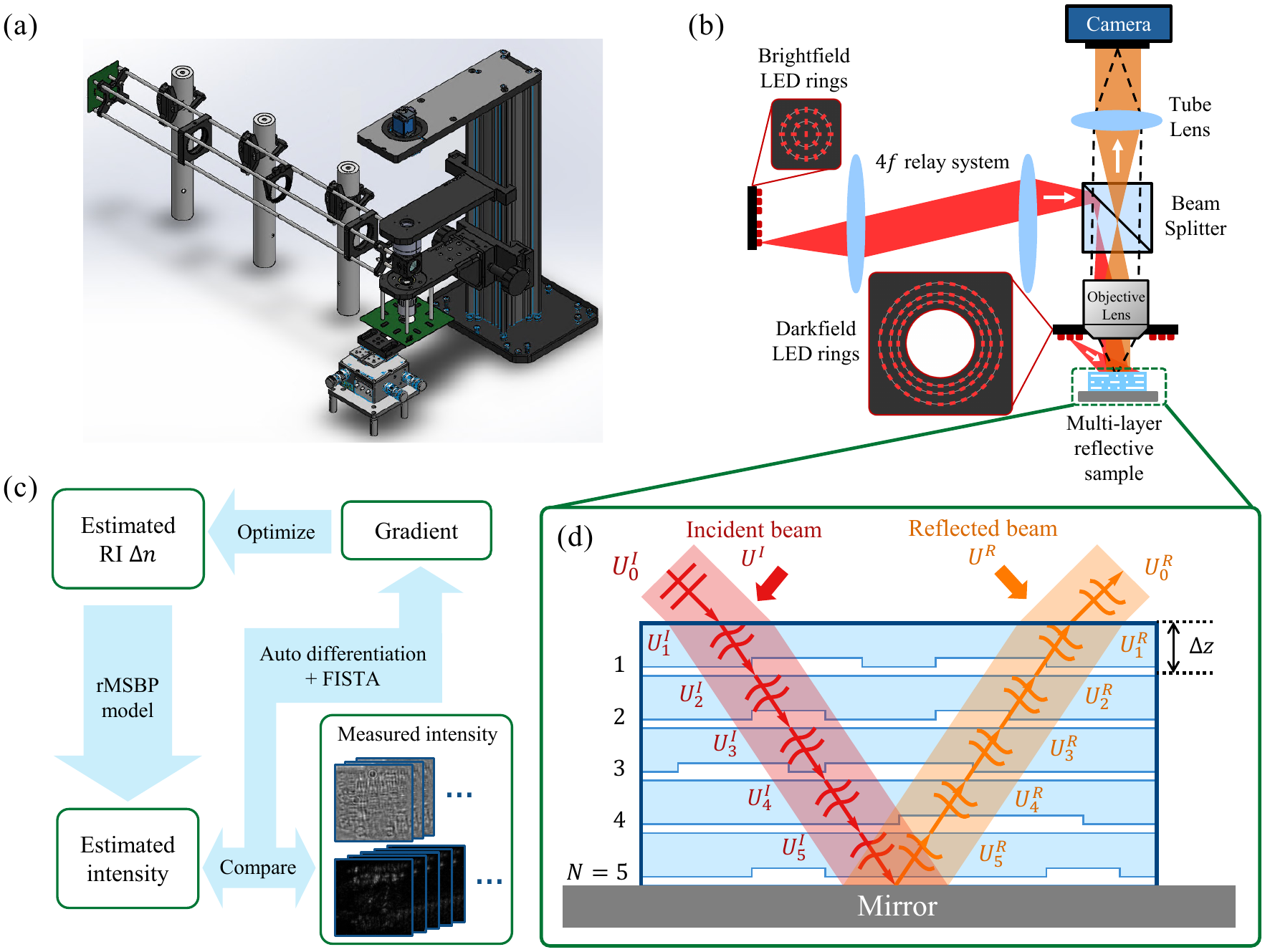}
\caption{
    Overview of the rMS-FPT System and Imaging Workflow.
    (a) 3D schematic of the rMS-FPT system.
    (b) Hardware setup.
    (c) Flowchart of iterative rMS-FPT reconstruction with rMSBP model.
    (d) Details of the rMSBP algorithm.}\label{fig:ridt_overview}
\end{figure*}

As the use of DT grows in semiconductor technology, the increasingly complex 3D architectures used in advanced devices present substantial challenges.
These architectures, essential for enhancing device performance, are difficult to characterize accurately using traditional methods.
Ensuring accurate measurements is crucial not only for optimizing production yields but also for accelerating research and development cycles~\cite{orji2018metrology}.

A primary challenge arises from the combination of strong scattering processes and substrate reflections in the metrology tasks.
Strong scattering processes occur due to high RI contrasts between the structures of interest and the surrounding materials.
Coupled with the complexity of 3D architectures and substrate reflections, these RI differences produce intricate light propagation phenomena~\cite{waterman1961multiple,bruning1971multiple}, including multiple scattering and shadowing effects.
Single-scattering approximations~\cite{wolf1969three,ling2018high}, commonly used in conventional DT, are insufficient to accurately model these interactions.
Multi-slice methods~\cite{kamilov2015learning,tian20153d, lim2019high, chen2020multi,zhu2022high} have emerged as computationally efficient alternatives, sequentially modeling light propagation through layered structures.
However, existing multi-slice approaches are primarily designed for transmission-mode DT and must be adapted to handle reflective substrates.

Pioneering works in reflection-mode DT~\cite{mudry2010mirror,zhang2013full} have shown promising results using the discrete dipole approximation (DDA), which models multiple scattering by discretizing the structure into a finite number of dipoles.
While DDA provides a more accurate representation of complex scattering, its computational cost scales significantly with the number of dipoles, limiting its efficiency for large structures or high-contrast materials.
On the other hand, single-scattering models in reflection-mode DT~\cite{matlock2020inverse,li2025transfer} are more computationally efficient but limited to imaging thin or weakly scattering samples.
Recently, the modified Born series (MBS) method~\cite{osnabrugge2016convergent} was introduced for reflection-mode DT~\cite{li2024reflection} and achieved significant improvements in efficiency and reconstruction accuracy.
However, it still has limited scalability for performing reconstructions on wide fields of view (FOV) and across large imaging depths.

While other metrology tools~\cite{schwenke2002optical}, such as scanning confocal microscopy or conventional intensity-only imaging, are commonly employed, they each entail inherent trade-offs.
Scanning-based methods are often too slow for wide-field inspection, whereas traditional intensity-only imaging lacks the quantitative phase sensitivity required to accurately reconstruct 3D refractive index (RI) distributions.

To address these limitations, we propose reflection-mode Multi-Slice Fourier Ptychographic Tomography (rMS-FPT), which integrates Fourier Ptychography (FP)~\cite{zheng2013wide, tian2014multiplexed, guo2015fpmMulti, lee2019reflectiveFPM, zheng2021review} with diffraction tomography in a reflection-mode configuration.
Our approach employs a reflection-mode multi-slice beam propagation method (rMSBP) to address the reflective substrate and the strong scattering overlay, enabling large-scale accurate 3D RI reconstruction.
As shown in Fig.~\ref{fig:ridt_overview}(d), the proposed rMSBP discretizes the sample into multiple layers and sequentially calculates their field perturbations.
To account for substrate reflections, we modify the standard multi-slice beam propagation method (MSBP) by inserting a reflection step at the depth of the reflective substrate, followed by the reverse propagation through the sample (See details in  \textbf{APPENDIX})

Furthermore, our rMS-FPT approach integrates brightfield (BF) and darkfield (DF) measurements, effectively enabling synthetic aperture reconstruction in 3D Fourier space and achieving 3D resolution enhancement from the angle-diversity measurements.
This extends 3D FP~\cite{tian20153d} to multiple-scattering samples on a reflective substrate.
By incorporating both BF and DF measurements, rMS-FPT surpasses the traditional BF resolution limit, akin to FP~\cite{zheng2013wide}.
Our previous work has demonstrated 2D reflection-mode FP~\cite{wang2023fourier} for topography measurement on thin highly reflective structures.
This work effectively extends the technique to complex multi-layered samples, achieving full 3D optical sectioning.
Unlike interferometric DT methods, rMS-FPT uses intensity-only measurement~\cite{tian20153d, chen2020multi, zhu2022high}, without requiring interferometric setups or mechanical scanning.
This simplifies hardware design, enhances robustness, and eliminates sensitivity to vibrations and calibration errors, making it well-suited for industrial applications.

We validate our rMS-FPT technique in both simulations and experiments.
In simulations, we test the technique on a four-layer phase resolution target placed on a perfect mirror.
The reconstruction results demonstrate that our technique achieves a lateral resolution of \qty{750}{\nm} with strong optical sectioning ability at a \qty{10}{\um} layer separation using a 0.28 NA imaging system.
Experimentally, we implement rMS-FPT using a reflection-mode LED array microscope~\cite{wang2023fourier}, as shown in Fig.~\ref{fig:ridt_overview}(b).
We image a two-layer resolution target sample and a multi-layer scattering sample.
The experimental reconstructions achieve the expected lateral resolution of $\lambda / 0.84$ for the two-layer resolution target, corresponding to a resolution enhancement factor of $1.5\times$ the incoherent diffraction limit of a traditional BF microscope.
For the multi-layer scattering sample, we achieve clear reconstructions in all layers, including overlapping beads, integrated circuit pattern, and resolution target, across a $\sim \qty{200}{\um}$ depth range and \qtyproduct{1.2 x 1.2}{\mm} lateral FOV.

Our results demonstrate that rMS-FPT provides a robust and practical solution for high-resolution, wide-field 3D metrology on reflective substrates, overcoming the limitations of existing DT methods.
The technique's ability to accurately reconstruct high-resolution 3D RI distributions on complex multi-layered samples across a wide FOV makes it well-suited for industrial metrology and inspection applications.

Our rMS-FPT strategy characterizes the 3D RI distribution of a reflective-substrate-supported samples from measured intensity images under diverse illumination angles, using the reflection-mode LED microscope, as shown in Fig.~\ref{fig:ridt_overview}.
The reconstruction process involves solving the inverse scattering problem, where the scattering process is modeled using our proposed rMSBP framework, shown in Fig.~\ref{fig:ridt_overview}(d).
The difference between the rMSBP prediction and the measured intensity formulates the loss function for the inverse problem.
As rMSBP is a non-linear forward model, we employ a gradient-based algorithm to iteratively update the estimated 3D RI distribution and minimize the loss function.
The details of the rMSBP algorithm are provided in the \textbf{APPENDIX} section.
With the rMS-FPT system and algorithm, we validate the performance of our rMS-FPT strategy in both simulations and experiments.

\section{Results}
\subsection*{rMS-FPT validation in simulation}

\begin{figure*}[ht]
  \centering
  \includegraphics[width=0.95\textwidth]{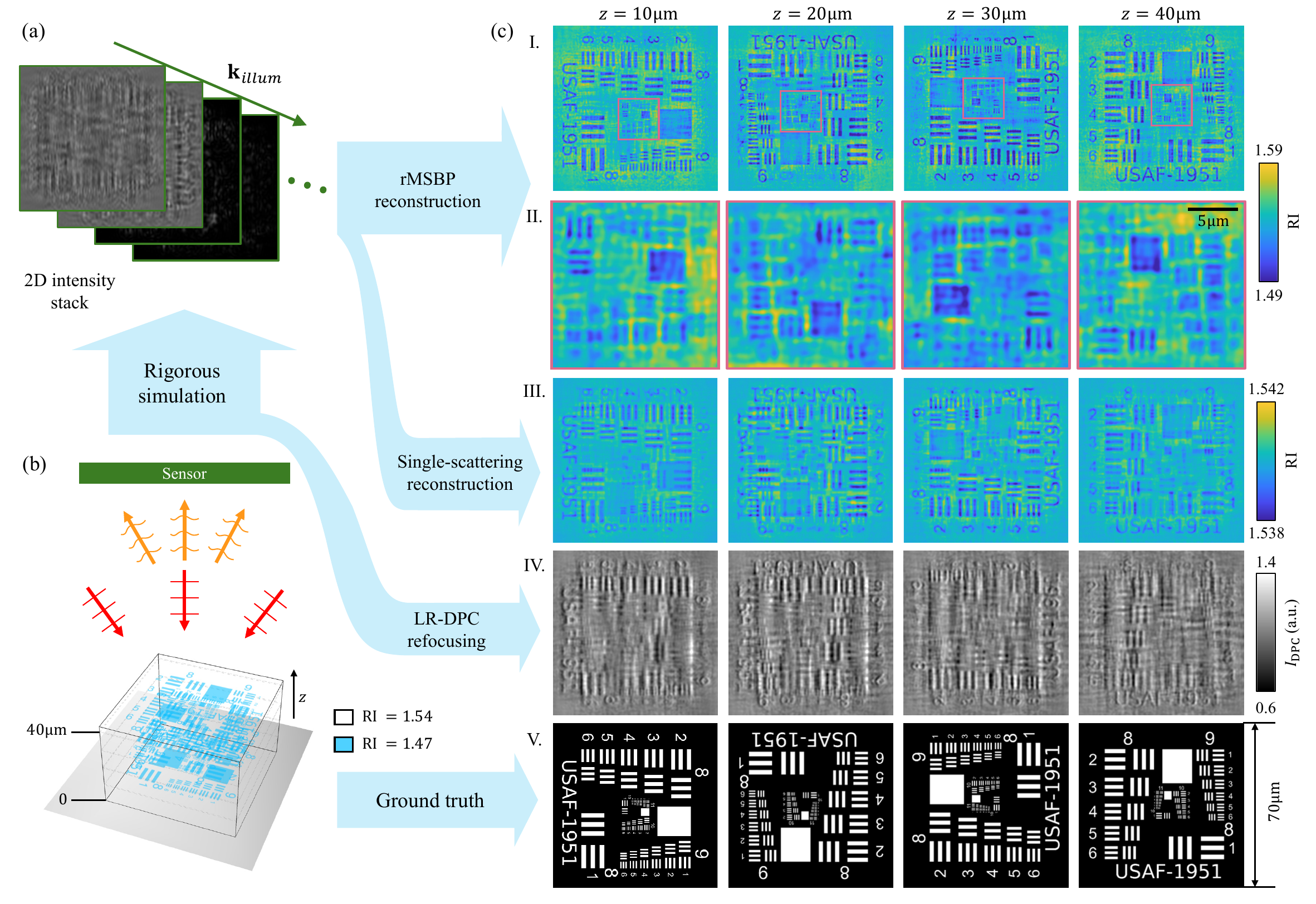}
  \caption{
    Performance evaluation of rMSBP on a simulated four-layer resolution test chart.
    (a) Generation of the 2D intensity stack using rigorous simulation based on the setup in (b).
    (b) Schematic of the four-layer 3D resolution target.
    (c) Comparison of reconstruction results at different depths (\qtylist{10;20;30;40}{\um}).
    (I) Reconstructed refractive index (RI) using rMSBP.
    (II) Magnified regions from (I).
    (III) Reconstructed RI using the single-scattering model.
    (IV) Reconstructed intensity using LR-DPC refocusing.
    (V) Ground truth RI. RI colorbars are shown for (I) and (III), intensity colorbar for (IV).
  }\label{fig:ridt_sim}
\end{figure*}

We first evaluate the performance of our rMS-FPT technique on simulated data.
In the simulation, the objective lens has an NA of 0.28.
The sample is illuminated by \qty{632}{\nm} monochromatic plane waves from 25 angles within the BF (incident NA$<$0.28) and 84 angles in the DF (incident NA$>$0.28) to generate angle-diverse intensity images.
The maximum NA of the DF incidence is 0.56.
These specifications are selected to balance resolution and FOV for industrial inspection applications.

We employ MBS as the rigorous scattering model to generate intensity images for the reconstruction algorithm validation, as established in our previous work~\cite{li2024reflection}.
The testing object is a four-layer 1951 USAF resolution test chart occupying a volume of \qtyproduct{70 x 70 x 40}{\um}, as shown in Fig.~\ref{fig:ridt_sim}(b).
The background RI is $n_0=1.54$, and the pattern in each layer has an RI of $n_1=1.47$ with a thickness of \qty{300}{\nm}.
The distance between adjacent layers is \qty{10}{\um}.
Each layer has a distinct pattern by rotating the original resolution target counterclockwise by \numlist[parse-numbers=false]{\ang{0};\ang{90};\ang{180};\ang{270}} from top to bottom, respectively (Fig.~\ref{fig:ridt_sim}(c)~(V)).
For this simulation, to test the lateral reconstruction fidelity, the reconstruction was performed by optimizing the RI only on the four known axial planes.
A detailed analysis of axial sectioning and crosstalk is presented in Supplementary Note II.

The reconstruction results are shown in Fig.~\ref{fig:ridt_sim}(c)~(I-II), where the test patterns in different layers are clearly resolved.
The reconstruction achieves a lateral resolution corresponding to a \qty{435}{\nm} line width (element 2, group 10), demonstrating a $2.6\times$ enhancement over the coherent diffraction limit ($\lambda/\mathrm{NA}_0$) set by single-LED illumination.
This approaches the theoretical $3.0\times$ enhancement (\qty{376}{\nm} limit) enabled by the combined brightfield and darkfield synthetic aperture.
We attribute this discrepancy to inter-layer crosstalk artifacts that arise when the \qty{10}{\um} layer spacing approaches the system’s theoretical axial resolution limit.
Based on the single-scattering model~\cite{li2025transfer}, we estimate the axial resolution to be approximately \qty{12.3}{\um}.
We have performed a more detailed investigation of this effect (see Supplementary Note II), which systematically analyzes the relationship between layer spacing and artifact severity, confirming this hypothesis.
This achieved resolution significantly surpasses the \qty{564}{\nm} limit of conventional incoherent brightfield imaging using the same 0.28 NA objective.
Our technique successfully resolves overlapping patterns with accurate phase recovery.
These results demonstrate the robustness and accuracy of our rMS-FPT strategy in reconstructing complex 3D RI distributions.

To highlight the advantages of rMS-FPT over physical scanning, we simulate intensity images focused on each layer and synthesize asymmetric illumination differential phase contrast (DPC) images along the left-right (LR) direction~\cite{tian20143d} in Fig.~\ref{fig:ridt_sim}(c)~(IV) (see \textbf{APPENDIX} for details).
While DPC offers a fast method for visualizing phase information, it requires physical focus scanning for 3D samples.
Moreover, its resolution is constrained by the BF diffraction limit, and its optical sectioning capability remains limited~\cite{chen20163d}.

To further highlight the importance of modeling multiple scattering in rMS-FPT, we compare reconstruction results obtained using our proposed rMSBP method with those from the single-scattering model~\cite{li2024reflection, li2025transfer}.
As shown in Fig.~\ref{fig:ridt_sim}(c)~(III), the lateral cross sections reconstructed using the single-scattering model exhibit significant crosstalk between layers, as expected.
On the other hand, the optical sectioning is significantly improved by the rMSBP method.

To further validate our modeling choice, we conducted a detailed comparison between the proposed rMSBP model and a reflection-mode multi-layer Born (rMLB) model, in which the MLB scattering and propagation operators introduced in~\cite{chen2020multi} are implemented within the same double-pass reflection geometry, replacing the BPM operators.
The results of this analysis are presented in Supplementary Note I.
This analysis confirms that the rMSBP model achieves an excellent balance of accuracy, numerical stability, and computational efficiency under our experimental conditions.
While the rMLB model provides a more rigorous treatment of non-paraxial effects and is advantageous for high-NA systems, its strong sensitivity to axial discretization ($\Delta z$) makes it less suitable for large-volume, high-throughput reconstructions.
In contrast, the rMSBP model remains robust even with coarse axial sampling and accurately reproduces the ground-truth intensity data, confirming its suitability for both forward and inverse modeling in rMS-FPT.

\subsection*{rMS-FPT validation in experiment}

\begin{figure*}[ht]
    \centering
    \includegraphics[width=0.85\textwidth]{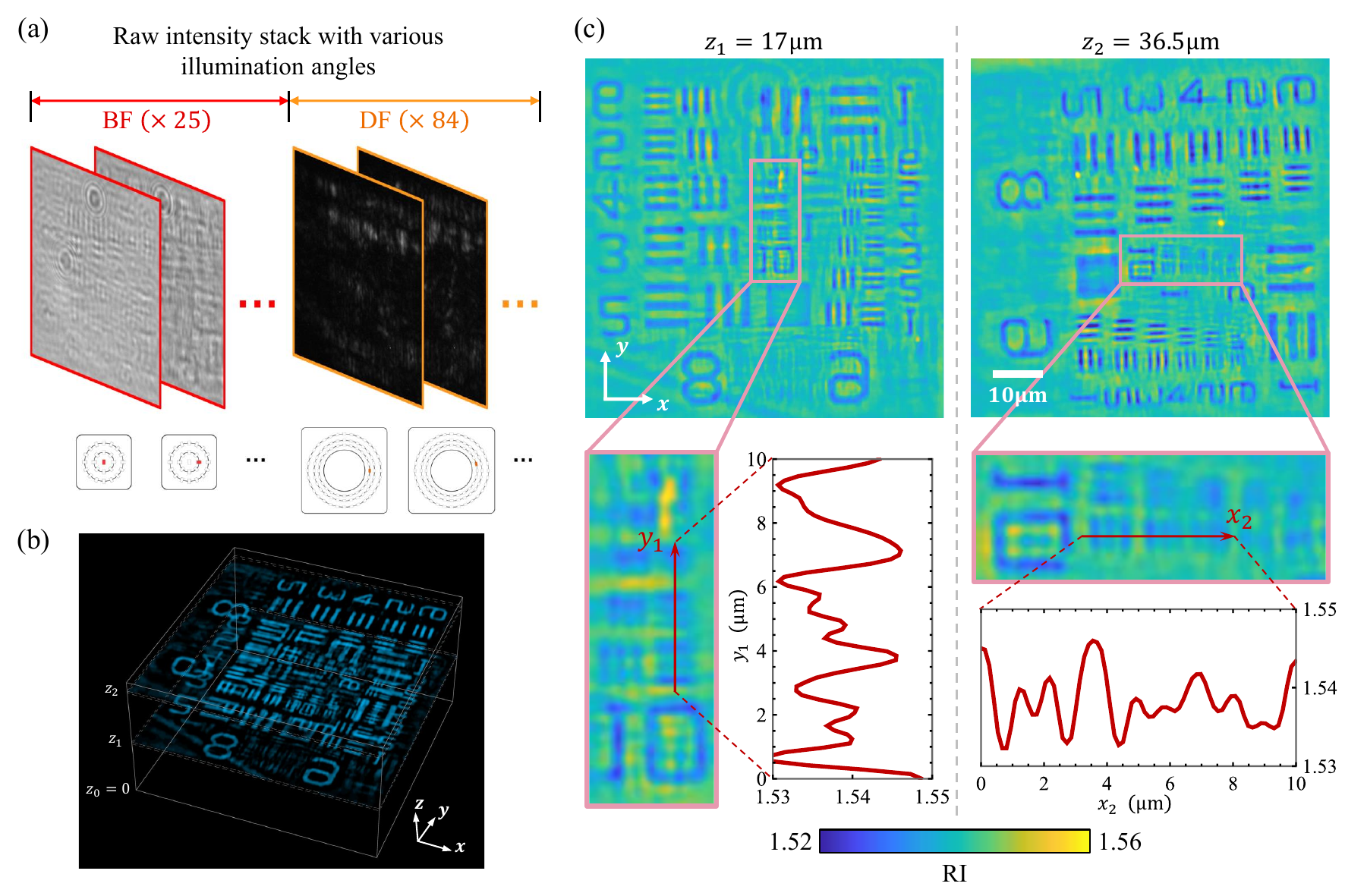}
    \caption{3D Reconstruction of a Dual-Layer Resolution Test Chart Using rMS-FPT\@.
      (a) Raw intensity measurements of the dual-layer resolution test chart along with  the corresponding LED illumination positions.
      The intensity of  DF images is enhanced  $50\times$ for better visualization.
      (b) 3D rendering of the reconstructed RI distribution,  where color opacity represents 
      RI values ranging from 1.53 to 1.54.
      (c) Reconstructed RI at depths of \qtylist{17;36.5}{\um}.
    }\label{fig:rFPT_exp}
\end{figure*}

To validate our simulation findings experimentally, we implemented rFPT using the reflection-mode LED microscope (Fig.~\ref{fig:ridt_overview}(b-c)), imaging a dual-layer phase resolution target on a silver mirror under conditions closely matching the simulation (see \textbf{Experimental setup} and \textbf{Sample Preparation}).
Raw intensity data (Fig.~\ref{fig:rFPT_exp}(a)) exhibit superimposed intensity patterns from the overlapping structures and reflections.
Applying the rMSBP algorithm yielded the reconstructed 3D RI distribution (Fig.~\ref{fig:rFPT_exp}(b)), clearly distinguishing the two layers axially at depths of \qtylist{17;36.5}{\um} over a \qtyproduct{140 x 140}{\um} FOV.
Key performance is assessed from the cross-sections (Fig.~\ref{fig:rFPT_exp}(c)), where element 2 of group 10 (\qty{435}{\nm} line width) is clearly resolved in both layers.
This achieved resolution corresponds to a $2.6\times$ enhancement over the single-LED illumination limit, matching the performance observed in the simulation (Fig.~\ref{fig:ridt_sim}).
This agreement between experiment and simulation confirms the robustness and practical applicability of our rMSBP-based rFPT approach for challenging reflection-mode imaging tasks.

\begin{figure*}[ht]
    \centering
    \includegraphics[width=0.97\textwidth]{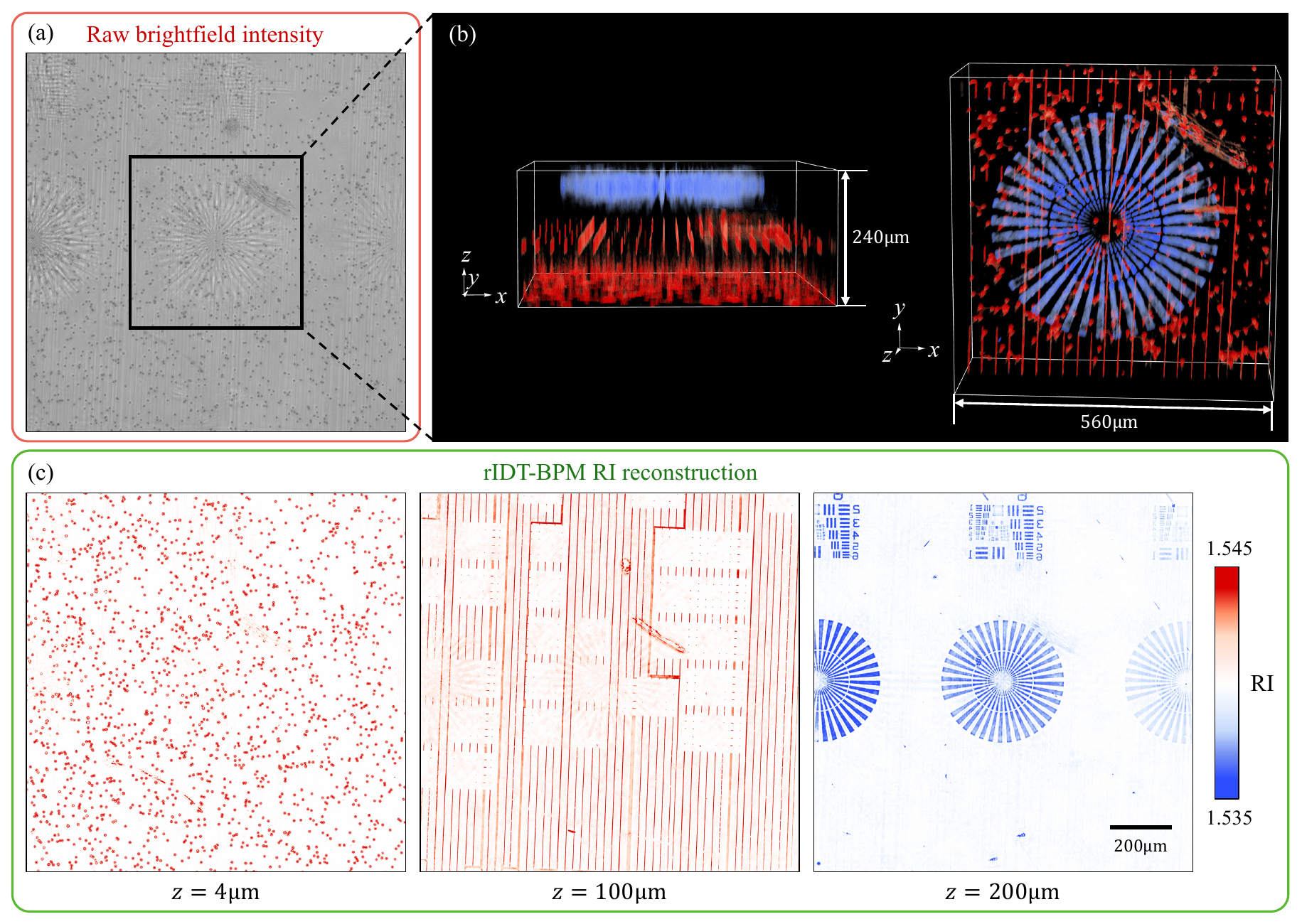}
    \caption{3D Reconstruction of a Multi-Layer Scattering Sample Using rMS-FPT\@.
      (a) Raw intensity measurement of the multi-layer scattering sample with all 25 BF LEDs simultaneously illuminated.
      (b) 3D rendering of the reconstructed RI distribution, with transparency applied to highlight RI variations from the background. The color mapping is consistent with (c).
      (c) $x$-$y$ cross sections of the reconstructed RI distribution at depths of \qtylist{4;100;200}{\um}.
    }\label{fig:rFPT_exp2}
\end{figure*}

Next, we apply rMS-FPT to a more complex multi-layer scattering sample, demonstrating its ability to rapidly reconstruct intricate 3D structures over a millimeter-scale FOV\@.
The sample consists of a phase resolution target on top, an integrated circuit (IC) pattern in the middle, and a beads layer at the bottom, all fabricated with photopolymer on a silver mirror (see Sample Preparation for details).
This design mimics the overlaid 3D structures encountered in semiconductor metrology and inspection.
Using the same experimental setup, the microscope captures a \qtyproduct{1.2 x 1.2}{\mm} FOV with a \qty{0.274}{\um} pixel size in object space.

For multi-layered samples thicker than the depth-of-field of the objective, the defocused layers spread scattered fields over a larger lateral region, causing highly unbalanced intensity responses across layers.
Combined with weak scattered signals, further attenuated by sample thickness, this leads to extremely uneven DF image intensities, challenging measurement due to limited camera's dynamic range and complicating rMSBP reconstruction.
Thus, to achieve practical high-speed imaging and reconstruction, only BF LEDs are used for measurement in this case.

Fig.~\ref{fig:rFPT_exp2}(a) shows a raw intensity image captured with all 25 BF LEDs simultaneously illuminated, focused at the mirror surface, to emulate a traditional incoherent brightfield microscope image.
For the rFPT reconstruction of this sample, these 25 LEDs were then used to provide angle-varied illuminations sequentially (see Table~\ref{tab:ridt_time} for details on the number of LEDs used in Exp.2).
In this measurement, all layers overlap laterally.
Since the RI contrast is weak relative to the background, only the in-focus beads layer is distinctly visible, while the out-of-focus layers manifest as faint, diffuse patterns due to diffraction and reflection from the substrate.

The reconstructed volume extends from the mirror surface to \qty{280}{\um} above, discretized into 70 slices with \qty{4}{\um} axial intervals and a \qty{0.274}{\um} lateral pixel size, matching the raw intensity images.
Due to the large-scale volume (\num{1.42e9} voxels) exceeding our GPU memory capacity, reconstruction is performed on 16 patches, each containing $1024 \times 1024$ pixels.
After reconstruction, patches are stitched together, and a morphological post-processing method~\cite{wang2023fourier} is applied to remove the background.
The reconstruction completes in 30 iterations, requiring \qty{382}{\s} per patch and a total of \qty{6113}{\s} using sequential processing.
Patch-based reconstruction can be parallelized on multi-GPU platforms to further accelerate processing.

Fig.~\ref{fig:rFPT_exp2}(b) presents two different views of a subregion of the 3D reconstructed RI distributions, the area of which is outlined in black in the intensity measurement in (a).
The three layers are clearly separated axially, with well-resolved RI contrasts.
The $x$-$y$ cross sections at $z = \qtylist{4;100;200}{\um}$ (corresponding to the beads layer, IC pattern layer, and resolution target layer) are shown in Fig.~\ref{fig:rFPT_exp2}(c).
The rMSBP reconstruction successfully extracts distinct RI distributions for each layer, with minimal crosstalk artifacts and clear visualization of the layered structure.

With a measurement duration of 1.56 seconds, this experiment demonstrates the capability of rMS-FPT to rapidly image and reconstruct multi-layer scattering RI distributions over a cubic-millimeter-scale volume on a reflective substrate.

\section{Conclusion}

We present a reflection-mode Fourier ptychographic diffraction tomography technique capable of characterizing the 3D RI distribution of highly scattering samples on a reflective substrate.
Integrated with a novel reflection-mode multi-slice beam propagation algorithm, our method provides a fast and accurate solution for modeling volumetric scattering processes in reflective geometries.
Through both simulations and experiments, we demonstrate rMS-FPT's ability to resolve complex overlapping structures.
Experimentally, our technique enables rapid data acquisition for a $1.20\times1.20\times\qty{0.28}{\mm^3}$ volume (1.6 seconds of measurement time), allowing for subsequent 3D reconstruction with sub-micron lateral resolution.

Our technique unlocks new possibilities for 3D characterization of novel devices in semiconductor, photonic devices, and micro-electromechanical systems.
For example, it provides a promising approach for wafer inspection, enabling the identification of subtle variations in material properties.
It also allows for deep defect detection crucial for ensuring device reliability.
The technique's versatility and adaptability highlight its potential to advance current metrology practices and address emerging challenges in nano-photonic and micro-fabrication technologies.

rMS-FPT bridges a critical gap by enabling wide-field, computationally efficient, and quantitative reconstruction of RI distributions.
However, due to the inherent ``missing cone'' problem in this imaging geometry~\cite{li2025transfer}, rMS-FPT is primarily sensitive to lateral RI variations.
This contrasts with optically sectioning techniques such as confocal microscopy~\cite{sheppard1990three}.
As purely axial structures contribute only to the phase of the unscattered DC component ($k_x = k_y = 0$) of the optical field, samples like perfectly flat, multilayer films provides little to no contrast in our setup implementation.
Instead, 3D optical sectioning is achieved by accurately reconstructing lateral features at their corresponding axial depths, as demonstrated experimentally in Figs.~\ref{fig:rFPT_exp} and\ref{fig:rFPT_exp2}.

We also acknowledge the inherent trade-off between the objective’s numerical aperture (NA) and overall system performance.
Although our approach is capable of modeling multiple scattering, samples with large refractive index mismatches can scatter light beyond the collection range of the 0.28~NA objective.
In principle, a higher-NA objective could capture this additional information; however, it introduces two major practical drawbacks.
First, the FOV would be substantially reduced, undermining the high-throughput, wide-area inspection capability central to this work.
Second, the resulting complex and non-paraxial scattering would demand more rigorous computational models to reconstruct information from the high-spatial-frequency regime.
Our system is therefore optimized for a practical metrology regime, achieving a balance between wide FOV, computational efficiency, and the capacity to accurately model multiple-scattering effects in multilayered structures.

Future work may focus on expanding rMS-FPT to handle a broader range of substrates, including rough surfaces, as surface roughness can degrade phase contrast and complicate imaging.
Additionally, optimizing the illumination scheme may mitigate sparse frequency support, particularly by the DF measurements, reducing artifacts and enhancing high-axial-resolution volumetric reconstructions.
Furthermore, incorporating polarization and backward scattering effects into the rMSBP model holds potential for capturing interfacial details, as strong backward scattering may reveal critical structural features at reflective boundaries~\cite{li2024reflection}.
Finally, incorporating advanced deep-learning techniques into the inverse-scattering process can further enhance reconstruction quality and computational efficiency, enabling faster and more accurate 3D imaging~\cite{liu2022recovery, wang2024neuph, matlock2023multiple}.

\section*{Data and Code Availability}
The reconstruction code used in this study is publicly available on GitHub at \url{https://github.com/bu-cisl/rMSBP-python}.
The experimental data supporting the findings of this study are available on \href{https://figshare.com/articles/dataset/Dataset_of_paper_i_Reflection-mode_Multi-slice_Fourier_Ptychographic_i_i_Tomography_i_/30573119}{figshare}.

\appendices
\section{}
\subsection{Experimental setup}
As shown in Fig.~\ref{fig:ridt_overview}(a), our rMS-FPT setup is built on Thorlabs Cerna\textsuperscript{\textregistered} microscope platform with 2 LED arrays.
The microscope is equipped with a $10\times$ objective lens ($\mathrm{NA}_0=0.28$, Mitutoyo Plan Apo Infinity Corrected Long WD) and a \qty{200}{\mm} focal length tube lens (Edmund Optics, MT-1).
Intensity images are captured by a monochrome CMOS camera (The Imaging Source, DMK 38UX 541) with a $4504 \times 4504$ (20.3 MP) pixel resolution sensor of \qty{2.74}{\um} pixel size (Sony, Pregious S IMX541).
We built our LED arrays with RGB LEDs (Kingbright APTF1616SEEZGQBDC) and the LED drivers (Texas Instruments, TLC5955) on custom-designed PCB boards.
The LED drivers are connected to a microcontroller (PJRC, Teensy 3.2) to control the intensity and color of each LED\@.
For the BF illumination, the LED array is placed on a side of the microscope, and we use a $4f$ system ($f_1=\qty{250}{\mm}$, $f_2=\qty{150}{\mm}$, Thorlabs ACT508-250-A, AC508-150-A) with a beam splitter (Thorlabs, CCM1-BS013) inserted between the objective lens and tube lens to relay the LED array to the back focal plane of the objective lens.
The $4f$ system is designed to demagnify the LED array by a 5:3 ratio, which roughly matches the pupil size of the \qty{20}{\mm} focal length objective.
This BF LED array consists of an LED at the center and 8 and 16 LEDs evenly distributed on two concentric rings.
For the DF illumination, we place the other LED array around the objective lens with 24, 28, and 32 LEDs evenly distributed on three concentric rings.
The DF LED array is placed at the height of the lowest surface of the objective lens, which is approximately \qty{34}{\mm} above the sample (same as the working distance of the objective lens).
The detailed geometrical parameters of the LED arrays are shown in Table~\ref{tab:led_array}.
After the measurement of the intensity images under diverse illumination angles, the incident angles are calibrated following the same method used in~\cite{eckert2016algorithmic,wang2023fourier}.

\begin{table*}[htb]
  \centering
  \caption{
    BF and DF LED arrays for the rMS-FPT system
  }\label{tab:led_array}
  \centering
  \begin{tabular}{ccccccc}
    \toprule
    & \multicolumn{3}{c}{BF LED Array} & \multicolumn{3}{c}{DF LED Array} \\[1ex]
    & Center & Ring 1 & Ring 2 & Ring 3 & Ring 4 & Ring 5 \\
    \midrule
    Designed NA & 0 & 0.130 & 0.251 & 0.420 & 0.490 & 0.560 \\
    Radius (\unit{\mm}) & 0 & 4.32 & 8.38 & 15.74 & 19.11 & 22.98 \\
    Demagnified Radius (\unit{\mm}) & 0 & 2.59 & 5.03 & -- & -- & -- \\
    Number of LEDs & 1 & 8 & 16 & 24 & 28 & 32 \\
    \bottomrule
  \end{tabular}
  \bigskip

\end{table*}

\subsection{Forward model}

The rMS-FPT imaging process is illustrated in Fig.~\ref{fig:ridt_overview}(b-d).
We model the object as a transparent volumetric medium with a perturbed RI distribution on a perfect mirror.
For each measurement, the object is illuminated by a quasi-monochromatic plane wave at a given angle of incidence.
The 3D rMS-FPT MSBP model simulates the scattering of the illumination field within the object.
Following the transmission-mode MSBP approach~\cite{tian20153d}, the scattering medium is discretized into slices parallel to the mirror surface.
The incident field propagates through and scatters from each slice, reflects off the mirror, and undergoes a second scattering process as it passes back through the slices in reverse order.
Mathematically, the scattering process can be described as follows.

First, we simulate the field propagation inside the transparent media:
\begin{equation}
  U_s^I(x,y)=e^{i\Delta\phi_s(x,y)} \iFxy{e^{i k_z \Delta z} \Fxy{U_{s-1}^I(x,y)}},
\end{equation}
where $\Fxys{\cdot}$ and $\iFxys{\cdot}$ denote the 2D Fourier and inverse Fourier transform on the XY plane, respectively,
the subscript $s$ is the slice index from the top at depth $z=z_s$,
$\Delta\phi_s(x,y)=k_0 \Delta z (n_s(x,y,z_s) - n_0)$ is the phase shift introduced by the $s$th slice,
$n_0$ is the background RI,
$k_z = \sqrt{k_0^2 n_0^2 - k_x^2 - k_y^2}$ is the axial component of the wave vector propagating along the optical axis,
$k_0$ is the center wavenumber in vacuum,
$k_x$ and $k_y$ are the lateral coordinates of the spatial frequencies, and
$U_s^I(x, y)$ is the optical field after scattered by the $s$th slice.

Second, upon reflection from a perfect mirror, the beam undergoes a phase change of $\pi$,
\begin{equation}
  U_N^R(x,y)=-U_N^I(x,y)
\end{equation}
where $U_N^I$ is the field scattered by the bottom slice closest to the mirror before reflection, and $U_N^R$ the field immediately after reflection.

Third, we simulate the reflected field propagating back through the sample again.
In this process, the axial component of the wave vector becomes $k'_z = -\sqrt{k_0^2 n_0^2 - k_x^2 - k_y^2}$, representing the wave propagates along the opposite direction along the optical axis:
\begin{equation}
  U_s^R(x, y)=\iFxy{ e^{-i k'_z \Delta z} \Fxy{e^{i\Delta \phi_{s+1}(x,y)} U_{s+1}^R(x,y)} }
\end{equation}
where $U_s^R$ is the reflected field scattered by the $s$th slice.
Note that the order of the scattering and propagation matrices is reversed after reflection --- the field is first scattered by each layer before propagating to the next layer, as shown in Fig.~\ref{fig:ridt_overview}(d).
The bottom layer, closest to the mirror, induces a phase shift both prior to and following reflection.
For a single-layer sample ($N=1$), this reduces to the 2D reflection-mode FP~\cite{wang2023fourier}.

Finally, we consider the field go through the microscope and captured by the camera with intensity $I$:
\begin{equation}
  U_{\mathrm{out}}(M x, M y) = \iFxy{P(k_x,k_y) e^{i k_z n_0 z_f} \Fxy{U_0^R(x,y)}}
\end{equation}
\begin{equation}
  I(M x, M y) = \left| U_{\mathrm{out}}(M x, M y) \right|^2
\end{equation}
where $M$ is the magnification, $z_f$ is the depth of the focal plane from the top slice, and $P(k_x,k_y)$ is the low-pass filtering by the pupil function on the frequency domain.
In practice, we ignore the aberration and use the ideal binary pupil function as
\begin{equation}
  P(k_x,k_y) = \begin{cases} 1, & \sqrt{k_x^2+k_y^2} \le k_0 \mathrm{NA}_0 \\
    0, & \sqrt{k_x^2+k_y^2} > k_0 \mathrm{NA}_0 \end{cases}
\end{equation}
Note that the pupil function can keep the unscattered component of the BF illumination and block that of the DF illumination, which gives a unified way to model all the illumination directions.

\subsection{3D reconstruction algorithm}
We reconstruct the 3D RI distribution by solving an inverse scattering problem using a gradient-based optimization algorithm.
The objective is to estimate the perturbed RI distribution that minimizes the discrepancy between the measured and simulated intensity images:
\begin{equation}
  \hat{\Delta n} = \argmin_{\Delta n} \sum_{l} \left\| \sqrt{I_l^m} - \sqrt{I_l^s(\Delta n)} \right\|_2^2 + \tau R_\mathrm{TV}(\Delta n).
\end{equation}
The first term ensures data fidelity, where $I_l^m$ represents the measured intensity for the $l$th illumination, and $I_l^s(\Delta n)$ is the corresponding simulated intensity.
The second term applies the 2D total variation (TV) regularization applied layer-wise, controlled by the regularization parameter $\tau$.
We choose the isotropic TV norm~\cite{kamilov2016optical} as
\begin{equation}
    R_\mathrm{TV}(\Delta n)=\sum_{x,y}\sqrt{{(\mathrm{D}_x \Delta n_{x,y})}^2 + {(\mathrm{D}_y \Delta n_{x,y})}^2},
\end{equation}
where $\mathrm{D}_x$ and $\mathrm{D}_y$ denote the finite difference operation along the respective direction.
We adopt a layer-wise 2D TV regularization instead of a full 3D TV for two main reasons.
First, it effectively enforces piecewise-constant structures in the lateral dimensions, consistent with the layer-by-layer nature of metrology samples.
Second, the system exhibits highly anisotropic 3D spatial frequency coverage, with substantially lower axial resolution.
Thus, applying a 2D regularizer independently to each slice provides a sufficiently accurate and computationally efficient approximation.

We use $\Delta n = 0$ as the initial guess and update the estimate iteratively by the principle of gradient descent.
Due to the complex-valued intermediate fields $U_s^I$ and $U_s^R$, we use the Wirtinger calculus with the auto-differentiation strategy to simplify the gradient calculation.
We define the gradient $\Grad$ of a complex function $\cvary(\cvarx)$ as $\Grad(\cvary,\cvarx)=\overline{\partial \cvary / \partial \cvarx+\partial\overline{\cvary}/\partial\cvarx}$.
For the data fidelity term $\Loss = \left\| \sqrt{I_l^m} - \sqrt{I_l^s(\Delta n)} \right\|_2^2$, we calculate the gradient of the complex fields with the chain rule.
The reflection field has gradient as
\begin{gather}
  \GradL{U_{out}} = 2 \ U_{out} \ \frac{\sqrt{I_l^s} - \sqrt{I_l^m}}{\sqrt{I_l^s}}\\
  \GradL{U_0^R} = \iFxy{
    \overline{P} e^{-i k_z n_0 z_f} \Fxys{\GradL{U_{out}}}
  }\\
  \GradL{U_{s+1}^R} = e^{-i\Delta \phi_{s+1}} \iFxy{
    e^{i k'_z \Delta z} \Fxys{\GradL{U_s^R}}
  }.
\end{gather}
This equation calculates the gradient for the field $U_{s+1}^R$ just before the $(s+1)^{th}$ slice by back-propagating the gradient $\GradL{U_s^R}$ through the free-space propagation from slice $s+1$ to $s$ and then through the phase modulation of slice $s+1$.
The incident field has gradient:
\begin{gather}
  \GradL{U_N^I} = -\GradL{U_N^R}\\
  \GradL{U_{s-1}^I} = \iFxy{
    e^{-i k_z \Delta z} \Fxys{e^{-i\Delta \phi_s} \GradL{U_s^I}}
  }.
\end{gather}
This equation calculates the gradient for the field $U_{s-1}^I$ just before the $s^{th}$ slice by back-propagating the gradient $\GradL{U_s^I}$ through the phase modulation of slice $s$ and then through the free-space propagation from slice $s-1$ to $s$.
Then, the gradient of $\Delta n$ can be calculated as:
\begin{gather}
  \GradL{\phi_s} = -i \left(\,\overline{U_s^I}\ \GradL{U_s^I} + \overline{U_s^R}\ \GradL{U_s^R}\right) \nonumber\\
  \label{eq:2grad} \GradL{\Delta n_s} = k_0 \Delta z \ \GradL{\phi_s}
\end{gather}
Eq.~\eqref{eq:2grad} shows that since each scattering slice scatters the beam twice during the forward and reflected propagation processes, both fields $U_s^I$ and $U_s^R$ contribute to the gradient of the RI distribution $\Delta n_s$.

We employ the Fast Iterative Shrinkage-Thresholding Algorithm (FISTA) to solve the optimization problem, using the gradient of the data-fidelity term and the proximal operator of the TV regularization term, following the approach in~\cite{kamilov2016optical}.

We use our high-performance auto-differentiation library~\cite{zhu2022high} to perform the computation on a server with an Nvidia Tesla V100 16G GPU\@.
The reconstruction times of the rMSBP algorithm for the simulation and experimental datasets are shown in Table~\ref{tab:ridt_time}.
\begin{table}[!ht]
  \centering
  \caption{
    Computational cost of the rMSBP reconstruction algorithm
  }\label{tab:ridt_time}
  \begin{tabular}{cccc}
    \toprule
    & Sim. & Exp.1 & Exp.2 \\
    \midrule
    FOV/\unit{\um^3} & \num{1.97e5} & \num{7.28e5} & \num{4.26e8} \\
    Pixels & $1024 \times 1024$ & $1024 \times 1024$ & $4504 \times 4504$ \\
    Layers & 4 & 2 & 70 \\
    LEDs & 109 & 109 & 25 \\
    Iterations & 50 & 20 & 30 \\
    Time/s & 94.7 & 27.4 & 6113 \\
    \bottomrule
  \end{tabular}
\end{table}

\subsection{Sample preparation}
The two-layer resolution target sample was fabricated on a flat silver mirror substrate with transparent photopolymer (Formlabs, Clear Resin, $\mathrm{RI}\approx1.54$).
The target patterns were first molded to two photopolymer films from a quantitative phase target (Benchmark Technologies, QPT) as master pattern.
Since the master pattern is a positive pattern, the molded photopolymer film became a negative pattern.
Two molded photopolymer films were then stacked to make the patterns overlap with each other and align at the center.
The negative patterns on the lower layer were filled with glycerin ($\mathrm{RI}=1.47$).
After the stacked layers being transferred to the mirror substrate, a cover glass was placed on top to package the sample.
Glycerin was also used to fill the negative patterns on the upper layer and the gap under the cover glass to remove the air, providing expected RI difference.

The large three-layer scattering sample was fabricated following a similar process.
First, we suspended polystyrene beads ($\mathrm{RI}=1.59$) in the liquid resin and cured as the thin beads layer.
Then, we added liquid resin on the cured beads layer and molded the integrated circuit pattern from a silicon wafer on the top.
Since this pattern is very dense, we used a mixture of isopropyl alcohol and liquid resin to fill it and covered with a polymer layer on the top, which was cured later to protect the pattern.
Finally, we molded a QPT layer on the two lower layers, filled glycerin on the top, and sealed the three-layer sample with a cover glass.

\ifCLASSOPTIONcaptionsoff
  \newpage
\fi



%

\bibliographystyle{IEEEtran}
\bibliography{ref}




%

\begin{IEEEbiographynophoto}{Jiabei Zhu}
received the B.S. degree in optics and optical engineering from University of Science and Technology of China (USTC), Hefei, China, in 2019, and the M.S. and the Ph.D. degree in electrical and computer engineering from Boston University in 2025. He is currently working as 3D imaging scientist at Nordson Test and Inspection. His research interests include scattering models, inverse problems, and automated optical inspection.
\end{IEEEbiographynophoto}

\begin{IEEEbiographynophoto}{Tongyu Li}
received the B.Sc. degree in physics from Tongji University, Shanghai, China, in 2018, and the Ph.D. degree in physics, from Fudan University, Shanghai, China, in 2023.
He currently is a postdoctoral scholar in Boston University, Boston MA, 02215, USA.
His recent search interests include computational imaging, scattering models, and inverse scattering problems.
\end{IEEEbiographynophoto}

\begin{IEEEbiographynophoto}{Hao Wang}
received the B.S. degree in optical information science and technology from University of Science and Technology of China USTC in 2016, the M.S. degree in optical engineering from the Shanghai Institute of Optics and Fine Mechanics, Chinese Academy of Sciences in 2019, and the Ph.D. degree in electrical and computer engineering from Boston University in 2024. His research interests broadly lie in optics and computational imaging.
\end{IEEEbiographynophoto}

\begin{IEEEbiographynophoto}{Yi Shen}
received the physics degree in physics from Nankai University, Tianjin, China, in 2023. He is currently pursuing the Ph.D. degree in electrical and computer engineering at Boston University, Boston, MA, USA, where he is a member of the Computational Imaging Systems Lab led by Prof. Lei Tian. He conducts research at the intersection of optics system and computational imaging.
\end{IEEEbiographynophoto}

\begin{IEEEbiographynophoto}{Guorong Hu}
received the B.Eng. degree from Jilin University, Changchun, China, in 2018, and the M.S. degree from the University of Michigan, Ann Arbor, MI, USA, in 2020, both in electrical and computer engineering. He is currently pursuing the Ph.D. degree in electrical and computer engineering at Boston University, Boston, MA, USA, where he is a member of the Computational Imaging Systems Lab led by Prof. Lei Tian. His research interests broadly lie in optics and computational imaging.
\end{IEEEbiographynophoto}

\begin{IEEEbiographynophoto}{Lei Tian}
is an Associate Professor and Director of the Computational Imaging Systems Group at Boston University. His research focuses on computational optics, with applications spanning biomedical imaging and semiconductor metrology. From 2013 to 2016, he was a postdoctoral associate in the Department of Electrical Engineering and Computer Sciences at the University of California, Berkeley. He received his Ph.D. (2013) and M.S. (2010) from the Massachusetts Institute of Technology. Dr. Tian is a Fellow of Optica. His honors include the 2025 Boston University Provost’s Scholar-Teacher of the Year Award, an NSF CAREER Award, the 2018 Boston University Dean’s Catalyst Award, the 2018 SPIE Fumio Okano Best 3D Paper Prize, the 2014 OSA Imaging Systems and Applications Best Paper Award, and the 2011 OSA Emil Wolf Outstanding Student Paper Prize.
\end{IEEEbiographynophoto}




\end{document}


%
\title{Supplementary Notes for Reflection-mode Multi-slice Fourier Ptychographic Tomography}
%
%
%

\author{Jiabei~Zhu,
        Tongyu~Li,
        Hao~Wang,
        Yi~Shen,
        Guorong~Hu,
        and~Lei~Tian
}

\maketitle

\IEEEpeerreviewmaketitle

\section{Comparison between the rMSBP and Multi-Layer Born Models}

\begin{figure*}[ht]
  \centering
  \includegraphics[width=0.9\textwidth]{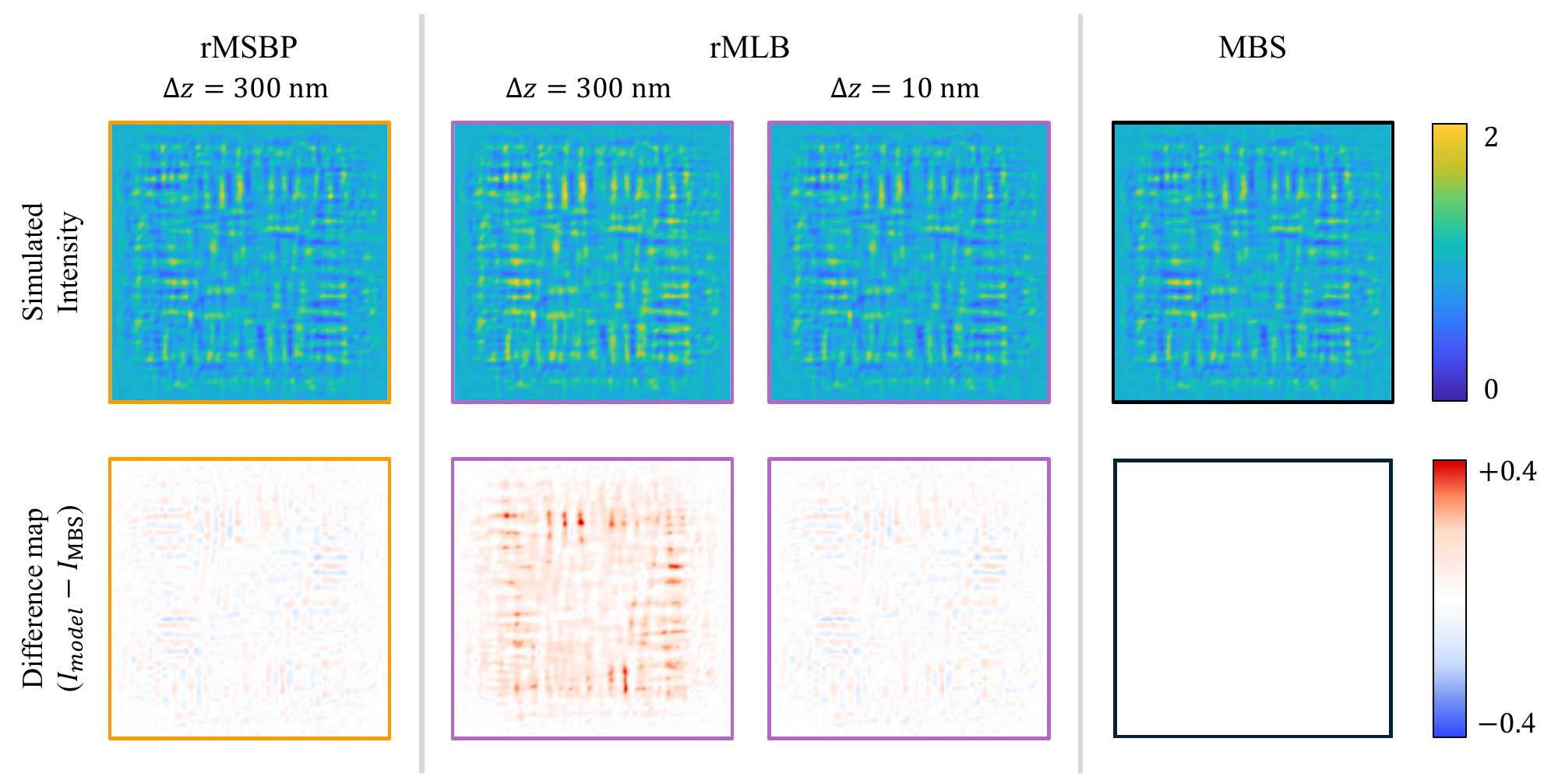}
  \caption{Qualitative comparison of forward models for a representative brightfield illumination (Ring 1, LED 4).
    (Top) Simulated intensity images from rMSBP ($\Delta z = \qty{300}{\nm}$) and rMLB ($\Delta z = \qty{300}{\nm}$ and $\qty{10}{\nm}$), compared to the rigorous MBS ground truth.
    (Bottom) The corresponding difference maps ($I_{model} - I_{\text{MBS}}$).
    The rMSBP model with sparse sampling shows minimal error, comparable to the densely sampled rMLB\@.
    The sparsely sampled rMLB, however, shows more prominent structural differences.
  }\label{fig:SI_fwd_img}
  \vspace{3pt}
  \includegraphics[width=0.95\textwidth]{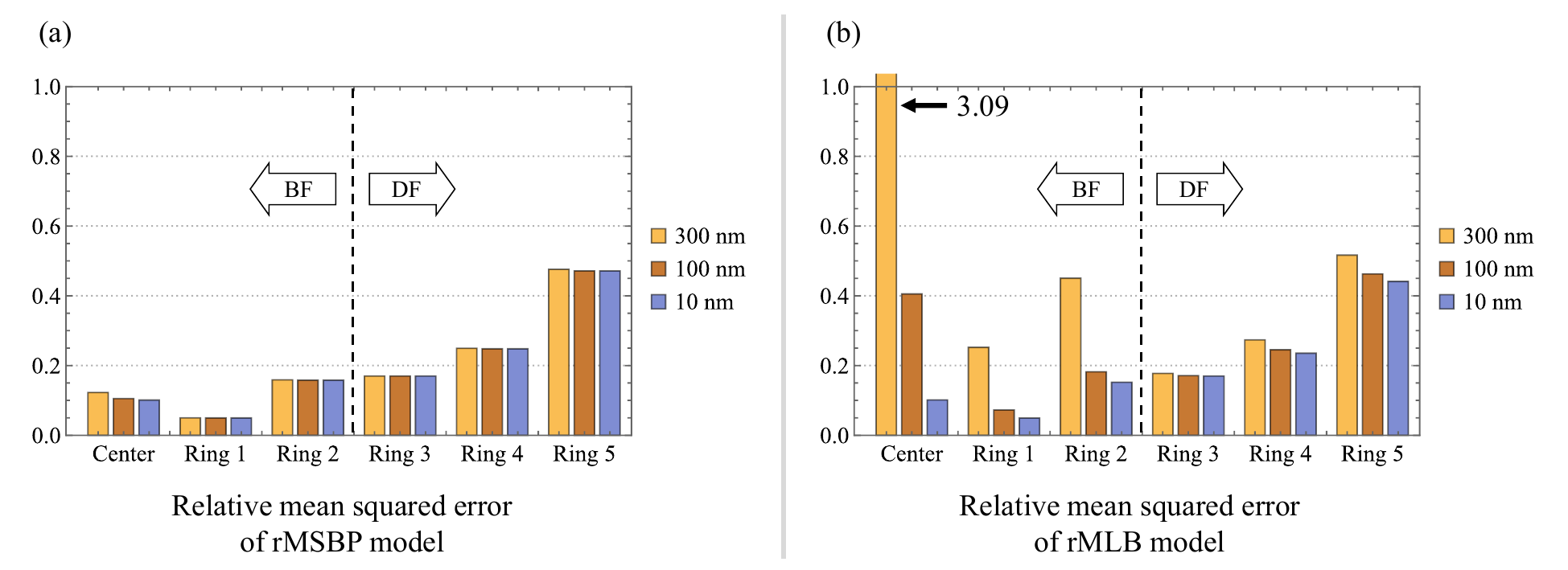}
  \vspace{3pt}
  \caption{Quantitative analysis of model error versus discretization step size across all BF and DF illumination rings.
    (a) Relative mean squared error (RMSE) for the rMSBP model.
    (b) Relative mean squared error for the rMLB model.
    The bars represent step sizes of \qty{300}{\nm} (yellow), \qty{100}{\nm} (brown), and \qty{10}{\nm} (blue).
    The RMSE is calculated as the mean squared error between the model and the rigorous MBS ground truth, normalized by the mean squared contrast of the ground truth images.
    The rMSBP model is stable and robust for all step sizes, while the rMLB model's accuracy is highly dependent on the step size.
  }\label{fig:SI_fwd_bar}
\end{figure*}

In the main text, we employ a reflection-mode multi-slice beam propagation (rMSBP) model.
As an alternative, we also propose the reflection-mode multi-layer Born (rMLB) model, which is a similar adaptation of the original transmission-mode multi-layer Born model~\cite{chen2020multi}.
The choice between these two models is primarily governed by two factors: their ability to account for non-paraxial scattering (determined by the system’s numerical aperture) and their robustness to axial discretization (i.e., the computational step size $\Delta z$).

Although the MLB framework provides a more rigorous treatment of non-paraxial effects and achieves higher accuracy in high-NA systems, our setup employs a low-NA ($\mathrm{NA}=0.28$) objective, where paraxial approximations remain valid.
Meanwhile, the robustness to axial discretization is particularly critical for computational efficiency in large-volume reconstructions.
We find that under our system conditions, the rMLB model exhibits strong sensitivity to $\Delta z$, whereas the rMSBP model demonstrates greater numerical stability and efficiency, making it a more practical choice for our application.

\subsection*{Simulation Methodology}

To compare the two models, we employed the same four-layer resolution target used in the main text simulations.
The sample consists of four \qty{300}{\nm}-thick 1951 USAF phase patterns with refractive indices of $n=1.54$ (background) and $n=1.47$ (pattern), corresponding to $\Delta n = -0.07$.
Each layer is separated by a \qty{10}{\um} gap, resulting in a total axial extent of approximately \qty{40}{\um} above a reflective substrate.
The rigorous Modified Born Series (MBS) model was used to generate the `ground truth'' intensity data for evaluation.

We then simulated the forward propagation using both the rMSBP and rMLB models.
This forward-model comparison alone provides a meaningful basis for assessment.
If a model fails to accurately reproduce measured intensities, it will inherently struggle in inverse reconstruction.
To evaluate sensitivity to axial discretization, we varied the step size $\Delta z$ from \qty{300}{\nm} (one step per layer) to \qty{100}{\nm} (three steps per layer) and \qty{10}{\nm} (30 steps per layer).

\subsection*{Results}

Fig.~\ref{fig:SI_fwd_img} presents a qualitative comparison under a representative brightfield illumination condition (Ring 1, LED 4).
The \qty{300}{\nm} step size corresponds to the discretization used in the main text reconstruction.
At this coarse step size, the rMSBP model output closely matches the MBS ground truth, with its difference map showing only weak, low-level residuals.
In contrast, the rMLB model at the same step size exhibits more noticeable structural discrepancies.
As the rMLB discretization is refined to \qty{10}{\nm}, both its reconstructed intensity and corresponding difference map become visually similar to those of the rMSBP model.

This behavior is quantified across all 109 illumination angles in Fig.~\ref{fig:SI_fwd_bar}.
As shown in Fig.~\ref{fig:SI_fwd_bar}(a), the relative mean-squared error (RMSE) for the rMSBP model remains consistently low across both brightfield (BF) and darkfield (DF) illumination rings, demonstrating strong robustness to variations in $\Delta z$.
Conversely, Fig.~\ref{fig:SI_fwd_bar}(b) shows that the rMLB model’s accuracy is highly sensitive to $\Delta z$.
At a coarse \qty{300}{\nm} step, the rMLB model exhibits substantial error, especially under on-axis and low-NA BF illumination.
Refining the step size causes this error to decrease markedly.
With sufficiently fine axial sampling (e.g., $\Delta z=\qty{10}{\nm}$), the rMLB model achieves accuracy comparable to that of the rMSBP model.

\subsection*{Conclusion}

This investigation demonstrates that the rMSBP model offers an optimal balance among accuracy, robustness, and computational efficiency for our application.
While the rMLB model remains valuable for high-NA systems where non-paraxial scattering dominates, its strong sensitivity to axial discretization renders it less practical for our high-throughput objectives, which benefit from a coarser computational grid.
Our results further confirm that the paraxial approximation underlying the rMSBP model remains highly accurate in this low-NA regime.

Moreover, the near-identical residual discrepancies observed in both the robust rMSBP model and the finely sampled rMLB model (Fig.~\ref{fig:SI_fwd_img}) suggest a common origin.
These errors likely arise from the intrinsic limitations of scalar, forward-scattering-only formulations --- the omission of vectorial and back-scattering effects.
Developing efficient strategies to incorporate these effects represents an important direction for future advancements of the rMS-FPT framework.


\section{Analysis of Inter-layer Crosstalk vs. Axial Resolution}

As discussed in the main text, the simulated reconstruction in Fig.~2, which used a \qty{10}{\um} layer spacing, showed minor crosstalk artifacts.
We hypothesized this was due to the \qty{10}{\um} spacing approaching the system's theoretical axial resolution limit (approximately \qty{12.3}{\um}).
To provide a quantitative investigation, we performed additional simulations to systematically analyze the relationship between layer spacing and artifact severity.

\subsection*{Simulation Methodology}
We used the same four-layer 1951 USAF resolution target described in the main text's simulation. We generated three separate "ground truth" datasets using the rigorous MBS-based forward model, each with a different axial spacing $h$ between the layers: \qty{6}{\um}, \qty{10}{\um}, and \qty{25}{\um}. The reconstruction for each dataset was then performed using our rMSBP algorithm.

\subsection*{Results and Analysis}

\begin{figure*}[ht]
  \centering
  \includegraphics[width=0.9\textwidth]{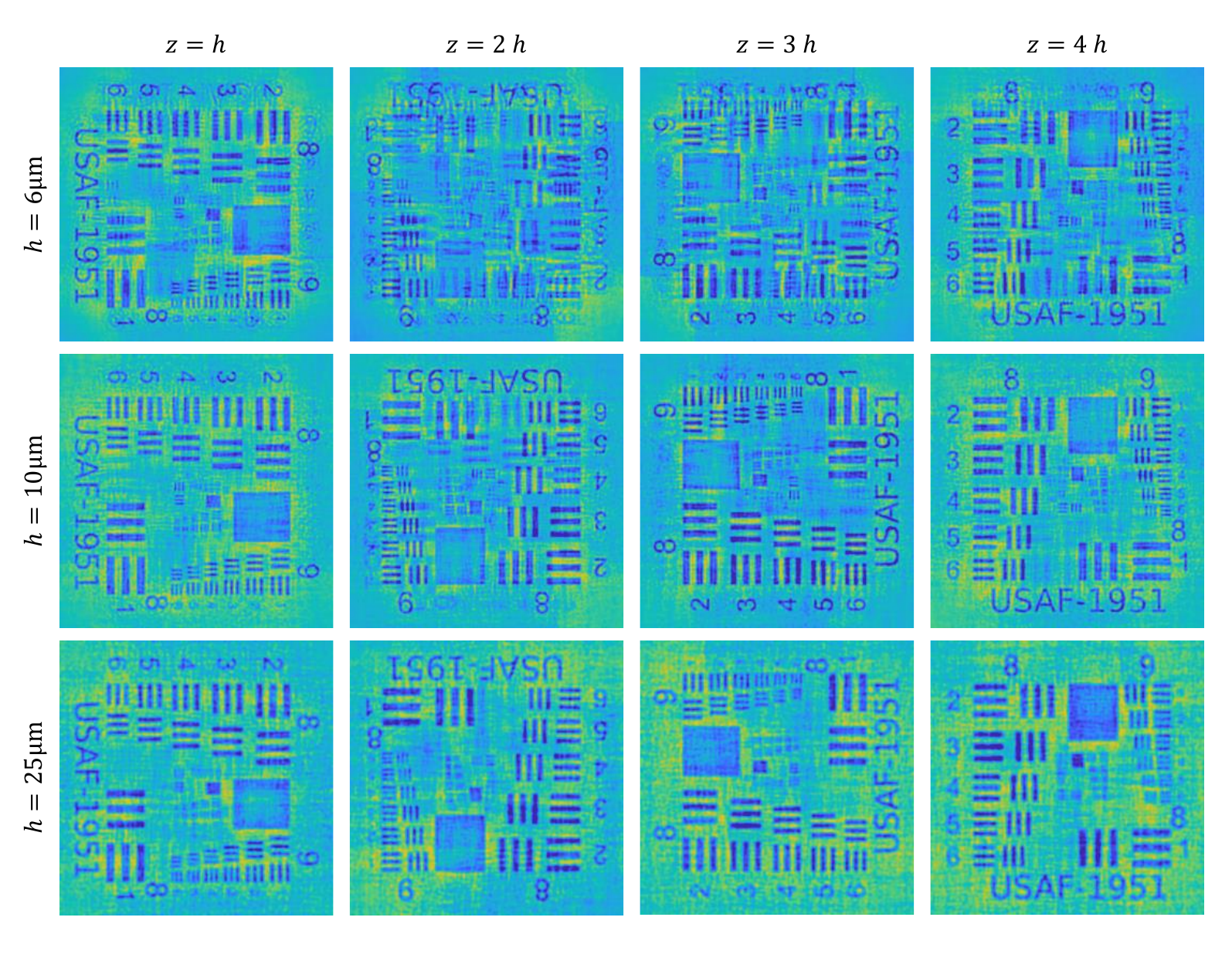}
   \caption{Crosstalk analysis versus layer spacing.
     Reconstructed $x$-$y$ cross-sections of the four simulated layers, shown at their respective depths $z=h, 2\,h, 3\,h, 4\,h$ (columns).
     Each row corresponds to a different simulation with a different axial gap $h$: \qty{6}{\um} (top row), \qty{10}{\um} (middle row), and \qty{25}{\um} (bottom row).
  }\label{fig:SI_crosstalk}
\end{figure*}

Fig.~\ref{fig:SI_crosstalk} presents the reconstruction results, clearly demonstrating the relationship between axial resolution and inter-layer crosstalk:

\begin{enumerate}
    \item \qty{6}{\um} gap (Top Row): At a spacing significantly below the system's approximate \qty{12.3}{\um} axial resolution limit, the algorithm cannot effectively separate the layers.
      This results in significant crosstalk artifacts, with features from different layers clearly bleeding into one another.

    \item \qty{10}{\um} gap (Middle Row): This is the case presented in the main paper.
      The spacing is around the limit of the axial resolution.
      As a result, we observe intermediate crosstalk, which is visible but does not prevent identification of the patterns.

    \item \qty{25}{\um} gap (Bottom Row): At a spacing well above the axial resolution limit, the reconstruction is clean, with minimal to no crosstalk artifacts.
      The layers are clearly and independently resolved, confirming the method's optical sectioning capability when the sample geometry permits.
\end{enumerate}

We note that for the \qty{25}{\um} gap, the overall sample thickness is considerably larger (approximately \qty{100}{\um} in total).
Such a long propagation path leads to substantial spreading and attenuation of the dark-field illumination signals, thereby slightly degrading the effective lateral resolution (e.g., features in group 10 become more difficult to resolve).
This behavior aligns with the general challenges of imaging thick specimens, as also discussed in the context of the multi-layer scattering sample (Fig.~4) in the main manuscript.

\subsection*{Conclusion}
This systematic analysis confirms that the crosstalk artifacts observed in the main paper are not a flaw of the rMSBP model.
Instead, they are a predictable physical consequence of the system's fundamental axial resolution limit. This study quantitatively verifies our system's optical sectioning capability and its limitations.

\ifCLASSOPTIONcaptionsoff
  \newpage
\fi



%

\bibliographystyle{IEEEtran}
\bibliography{ref}




%



